# On A New Formulation of Micro-phenomena: Basic Principles, Stationary Fields And Beyond

Afshin Shafiee[1]
Research Group On Foundations of Quantum Theory and Information,
Department of Chemistry, Sharif University of Technology,
P.O.Box 11365-9516, Tehran, Iran.

## Abstract

In a series of essays, beginning with this article, we are going to develop a new formulation of micro-phenomena based on the principles of reality and causality. The new theory provides with us a new depiction of micro-phenomena assuming an unified concept of information, matter and energy. So, we suppose that in a definite micro-physical context (including other interacting particles), each particle is enfolded by a probability field whose existence is contingent upon the existence of the particle, but it can locally affect the physical status of the particle in a context-dependent manner. The dynamics of the whole particle-field system obeys deterministic equations in a manner that when the particle is subjected to a conservative force, the field also experiences a conservative complex force which its form is determined by the dynamics of particle. So, the field is endowed with a given amount of energy, but its value is contingent upon the physical conditions the particle is subjected to. Based on the energy balance of the particle and its associated field, we argue why the field has a probabilistic objective nature. In such a way, the basic elements of this new formulation, its application for some stationary states and its nonlinear generalization for conservative systems are discussed here.

## 1 Introduction

There exists a micro-phyisical world independent of our observations obeying deterministic causal rules. This is the first principle of a new formalism which we are going to elucidate its scope and extent in a series of papers beginning with constructing the foundations of it in the first essay. In this new approach, while we presume the principles of realism and causality based on the classic-like equations of motion, we explore the meaning of the wave function to explain why the its form (according to Born postulate) determines the probability density of finding a *particle* somewhere in space. Then we give a clear depiction for some weird quantum phenomena such as the tunneling effect [1], double-slit experiment [2] and the so-called Einstein, Podolsky and Rosen (EPR) thought experiment [1, 3]. We show also how the current Schrödinger equations can be extended to a more general form including nonlinear terms. So, the new theory presents new predictions which can be checked in practice.

From a more fundamental point of view, this theory provides with us a new formulation of quantum phenomena based on an unified concept of information, matter and energy. Here, we will see that the micro-particle can be viewed as an energy source which shares some of its energy with its surrounding, i.e., the space where the particle can be likely found within there after a suitable measurement is performed. This corresponds exactly to the meaning of a spatial probability distribution which is endowed with energy here. So, we suppose that there is a field associated with a particle which form together a unit entity called particle-field. The field has a function-dependency which determines the spatial distribution of the entire system, i.e., it determines the probability of *finding* the particle within a definite interval of space, when one measures its position. Moreover, both the particle and the field satisfy deterministic equations of motion, but the field has no independent identity without the existence of particle. Alike the particle, the field also experiences some definite forms of force in space. As a result, the field is also endowed with energy which its value is contingent upon the physical conditions the particle is subjected to. We are not able to see an independent particle directly without any intervention. Accordingly, a particle-field is indeed a particle which its nature differs from a classical particle because the particle shares some of its energy with its surrounding space. So, in micro-incidents, one cannot discern particle's properties or field's attributes individually in practice, but only their entire properties are discernible.

---

[1]E-mail: shafiee@sharif.edu

The matter (characterized by the existence of a particle), the energy (attributed to the whole particle-field), and information (by which we gain knowledge about the possible locations a particle can be found) are all unified concepts. This is somehow similar to the special relativity theory in which the concepts of matter and energy are joined into one concept in the relativistic domain. Consequently, the probability distributions have an *objective* character here, because they are endowed with energy. In classical domain, however, energy is a characteristic feature of the particle alone, so the probability distributions are not energetic fields, but only mathematical functions enabling us to make some predictions about the possibility of observing a particle in a given space proportional to the time it spends there. Such a significant assumption about the objective character of information which is intertwined with the whole physical nature of the system has undoubtedly great physical and metaphysical consequences which needs substantial attention. What we are doing here is just opening a new door for exploring micro-phenomena in an innovative fashion.

The structure of this paper is as the following. In section 1, the basic elements of the new theory is first described for a one-particle one-dimensional micro-system (for both time-independent and time dependent events) and then will be generalized to include many-body systems. In section 2, we first reexamine in detail the physical aspects of two elementary stationary problems, i.e., the translation and vibration of a particle in a causal deterministic approach. Then, the stationary states of the hydrogen atom is reconsidered. In section 3, the possible generalization of the time-independent Schrödinger equation to a non-linear equation is discussed and the general form of the translational energy levels for a particle in a one-dimensional box is given by solving the corresponding non-linear equation. We then summarize our results in section 4.

## 2  Basic elements

### 2.1  General traits

For a one-dimensional, one-particle micro-system, three physical entities are introduced:

**1**· A particle with mass $m$ and position $x(t)$ which its dynamics is given by the Newton's second law:

$$m\frac{d^2x(t)}{dt^2} = f_P \tag{1}$$

where $f_P$ is the force defined for the particle. For a conservative force, the particle possesses a conserved energy $E_P = V_P + K_P$, where $K_P = \frac{p_P^2}{2m}$ is the kinetic energy and $p_P$ is the linear momentum of particle.

2· Associated with the particle, there is a field denoted by $X(x(t), t)$ with velocity $v_F = \left|\frac{dX}{dt}\right| = |\dot{X}|$ along the positive direction of $x$, where

$$\dot{X} = \left(\frac{\partial X}{\partial x}\right)v_P + \left(\frac{\partial X}{\partial t}\right) \tag{2}$$

and $v_P$ is the velocity of particle along the same direction. The field has a dimension of length and similar to the particle, we assume that it obeys a Newton-like dynamic too,

$$m\frac{d|\dot{X}|}{dt} = f_F \tag{3}$$

where $f_F$ is the force the field is subjected to. If the particle is subjected to a conservative force $f_P$, we will consider $\chi = \chi(x(t))$. Then, one can show that

$$v_F = \left|\left(\frac{d\chi}{dx}\right)\right|v_P = |\chi'|v_P \tag{4}$$

and

$$f_F = mv_P^2 \frac{d|\chi'|}{dx} + f_P|\chi'| \tag{5}$$

As is obvious in relations (1) to (5), there is no interaction between the particle and its associated field. The force $f_F$ only includes the force of particle $f_P$, but encompasses no interaction force between the particle and field. Hence, from a physical point of view, the field $X$ merely enfolds the particle. It experiences its own mechanical-like force introduced as $f_F$ in (3), although the presence of particle is essential for defining the force of field. If there is no particle, there will be no associated field too. The existence of field depends on the existence of particle, but the opposite is not true. For, $X$ is a function of particle's position (denoted by $x(t)$ in (1)), and not vice versa.

For a conservative field subjected to the force $f_F$ in (5), one can define the energy $E_F = V_F + K_F$ where $K_F = \frac{1}{2}mv_F^2 = K_p|\chi'|^2$. The kinetic energy of field includes the kinetic energy of particle. Here, one cannot separate the meaning of $K_F$ from $K_P$.

In the quantum domain, both the quantities $E_P$ and $E_F$ are hidden (not practically discernible), but the total energy $E = E_P + E_F$ is an observable property. Now, let us write this equation as the following:

$$E = V_P + \left(E_F + \frac{p_P^2}{2m}\right) = V_P + \frac{p^2}{2m} \tag{6}$$

where $\frac{p^2}{2m} = \left(E_F + \frac{p_P^2}{2m}\right)$. As is apparent in relation (6), $\frac{p^2}{2m}$ has not essentially the meaning of a kinetic energy, because $E_F$ might be negative and greater than $\frac{p_P^2}{2m}$ in magnitude, so $p$ would be imaginary.[2] Nevertheless, for all situations where $\left(E_F + \frac{p_P^2}{2m}\right)$ is positive, $p$ can be regarded as a kind of linear momentum along the positive direction of $x$ in a one-dimensional problem. In these situations, the value of $p$ is generally hidden (for, $E_F$ is hidden), except for the cases that $V_P$ (and so $f_P$) is zero everywhere, so that $p^2$ can be obtained directly from the value of $E$ in (6). For the situations in which the second term in (6) is positive, $p$ has a double feature: it includes a field aspect introduced by $E_F$ and a particle aspect characterized by $K_P$. Due to this dual character, in such situations we postulate that $p$ should satisfy the so-called de Broglie relation $p = \frac{h}{\lambda} = \hbar k$, where $\lambda$ is the wavelength and $k = \frac{2\pi}{\lambda}$. It should be noted that de Broglie momentum is an observable, but its value is discernible in practice only when the particle is subjected to no forces whatever (i.e., for a free particle). In this situation, $f_P = 0$ and it is expected that the field oscillates as a requirement the physical meaning of $k$ imposes. One can see how $f_P$ fullfills this condition in relation (9) below.

If $E_F$ becomes zero under some definite circumstances, then $E$ in (6) will be equal to $E_P$. Consequently, we will have a field without energy and all classical features will appear afterwards. That is, we will have merely a particle not associated with a field. Hence, we consider $E_F \to 0$ as the classical limit. The implications and consequences of the classical limit will be explored in different situations subsequently.

Unfortunately, the form of the force $f_F$ in (5) is complicated and unknown *a priori*, so it is not possible to obtain $\chi$ from (3) or (5). Accordingly, we postulate that for stationary states which the energy is conserved, the form of $\chi(x(t))$ could be obtained from the time-independent Schrödinger equation:

$$\chi'' = -k^2 \chi \qquad (7)$$

where $k^2 = \frac{p^2}{\hbar^2} = \frac{2m}{\hbar^2}(E - V_P)$ and $\chi'' = \frac{d^2 \chi(x)}{dx^2}$.

According to the field $\chi(x(t))$, we define a quantum mechanical wave function (denoted by $\psi$) that satisfies the time-independent Schrödinger equation (7) too:

$$\psi(x(t)) = \frac{\mathcal{N}\chi(x(t))}{A} \qquad (8)$$

where $\mathcal{N}$ is a constant and $A$ is a real (not constant) parameter with dimension of length which is independent of $x$ and $t$. The physical significance of both parameters will be cleared later. A similar relation as (8) can be written for a time-dependent field $X(x,t)$ and its corresponding wave function $\Psi(x,t)$.

The general form of $\chi$ can be written as $Af(x(t))$, so that $\psi(x(t)) = \mathcal{N}f(x(t))$. As a result, $\chi$ has a function-dependency (denoted by $f(x(t))$) which is in common with $\psi$, but contains a factor $A$ with different values in various situations. Thus, in some circumstances (e.g., at the classical limit), $A$ may approach zero (and so $\chi$), but $\psi$ remains definite and non-zero.

---

[2]This indeed happens in the tunneling effect which we will survey it in [1].

This is an important point which shows the crucial difference between $\chi$ and $\psi$.

For stationary states in which $\psi$ (or $\chi$) is a real function, one can rewrite relation (5) as

$$f_F = -m\bar{\omega}^2 \chi + f_P \chi' \tag{9}$$

in which we have used (7) and $\bar{\omega}^2 = v_P^2 k^2$. This shows that when $\chi$ is real and relying on $x(t)$ only, the field -at least- partly experiences an oscillating-like force. Albeit, $\bar{\omega}$ depends on $x(t)$ and the first term in (9) does not actually describe an oscillating force, but having the same form as it. If one generalizes this term to contain anharmonic terms (e.g., including the terms $\chi^3$, etc.), it will be possible to derive a non-linear Schrödinger equation. We will return to this problem in section 3.

**3·** Neither the particle, nor the field representations alone are adequate for explaining the behavior of a micro-system comprehensively. What really gives us a thorough understanding of the nature of a quantum system is a holistic depiction of both particle and its associated field which we call it a *particle-field* (PF) system. The kinetic energy of a PF system is proportional to $K_P + K_F$, but its total energy is the same as $E$ in (6).

Let us define the kinetic energy of a PF as $K_{PF} = \frac{1}{2} m \dot{q}^2$, where $q$ denotes the position of the PF and $\dot{q}$ is its velocity. Then, it is legitimate to suppose that $K_{PF} \propto K_P + K_F$, or

$$\dot{q}^2 = g_{PF}^2 \left( \dot{x}^2 + |\dot{X}|^2 \right) \tag{10}$$

where $g_{PF}$ is a proportionality factor and $\dot{x} = v_P$. As we will see later, for many problems this factor is equal to one, but the non-oneness of its value in general is crucial in some other problems (see, e.g., section 3.1).

The above relation can be rewritten as a geometric relation in Euclidean space:

$$dq^2 = g_{PF}(dx^2 + |dX|^2) \tag{11}$$

From this relation, one can obtain the trajectories of a PF system:

$$q(x,t) = g_{PF} \int dx \sqrt{\left(1 + \left|\frac{dX(x,t)}{dx}\right|^2\right)} \tag{12}$$

The relation (11) shows that while we expect the particle moves along the infinitesimal displacement $dx$ in the $x$-direction, the displacement of the whole system is equal to $dq$, not the same as $dx$. The difference here is due to the existence of the associated field which adds a new term, in addition to the direction the particle goes along. Hence, the PF system indeed keeps going through an *integrated* path determined by the whole action of the particle and its associated field.[3]

Using the relation (12), one can obtain the finite displacement $q$ of a PF system in terms of particle's location $x(t)$ and time $t$, when the field $X(x,t)$ is known. Then, if the form of dependence of $x$ to $t$ is also known for a given physical problem, it is possible to write $q$ totally in terms of $t$. For stationary states, however, $q = q(x(t))$ and there is no explicit time-dependency. Therefore, one can see that the time variable could be kept concealed in trajectory equations, so that the spatial direction $x$ would be sufficient for illustrating the behavior of $q$.

---

[3] Two adjacent points which satisfy the relation $ds^2 = c^2 dt^2 - dq^2$ ($c$ is the velocity of light) can be connected by a light signal. This introduces a unified concept of spacetime in micro-world which could be considered as a geometric relation for other applications.

The dynamics of the PF system can be also described according to a Newtonian equation. So, we have:

$$m\frac{d^2 q}{dt^2} = f_{PF} \tag{13}$$

where $f_{PF}$ is the force the PF system is subjected to. Using the relations (4) and (10), one can get a relation for $f_{PF}$ in the stationary states (i.e., the states in which $X = \chi(x(t))$):

$$f_{PF} = g_{PF}\left[f_P(1+\chi'^2)^{1/2} + \frac{m\dot{x}^2 \chi' \chi''}{(1+\chi'^2)^{1/2}}\right] \tag{14}$$

If $f_{PF}$ is conserved, the energy of the PF system (which is the same energy obtained by solving the equation (7)) can be expressed as:

$$E = E_P + E_F = V_{PF} + K_{PF} = V_{PF} + g_{PF}^2 K_P (1 + [\chi'(x(t))]^2) \tag{15}$$

where $V_{PF}$ is the potential of the PF system.

From the above relation, it is possible to find an expression for $V_{PF}$:

$$V_{PF} = g_{PF}^2 (V_P + V_F) + E(1 - g_{PF}^2) \tag{16}$$

for which it is straightforward to show that $\frac{\partial V_{PF}}{\partial q} = -f_{PF}$.

Using the relation (10), we can show that in general,

$$K_{PF} = g_{PF}^2 K_P \left(1 + \left|\frac{dX(x,t)}{dx}\right|^2\right) \tag{17}$$

For simplicity, hereafter, we take $g_{PF} = 1$, except for the cases which needs more elaboration.

The relation (17) which demonstrates the significant role of $K_P$ in the definition of kinetic energy of the whole system, also helps us to interpret the meaning of $X$ (and so $\Psi$) clearly. To reach a coherent meaning of $\chi$, one should note that whenever $K_{PF}$ is minimized at a given location, one can find the enfolded particle more likely there.

For save of clarity, however, let us focus for now on the case of stationary states with real $\chi$s. The generalization to other forms of the field is then straightforward. The reason is that, mathematically, any kind of $X$ (or $\Psi$) can be written as a superposition of the appropriate real stationary fields which form a complete set of solutions in equation (7). So, the same interpretation as for real $\chi$s can be inferred for them too.

For a real stationary field, one can write relation (17) as:

$$K_{PF} = K_P(1 + \chi'^2) \tag{18}$$

Taking into account a typical location $x = x^*$, three extremum statuses are in order:

1) $K_P|_{x=x^*} = 0$. This situation happens when particle has a varying velocity. Usually, in locations where this condition is satisfied, the probability of finding the particle is maximum. However, there is (at least) one exception. Consider, e.g., a situation in which the energy is conserved for a PF system. As was stated earlier in relation (6), we can define $E = V_P + \frac{p^2}{2m}$ where $\frac{p^2}{2m} = (E_F + K_P)$ and $p = \frac{h}{\lambda} = \hbar k$. In this situation, if $K_P = 0$, then $p = \sqrt{2mE_F}$. Now, let us assume that there is a stationary state for which $E_F < 0$, but $(E_F + K_P)$ is positive at definite places like $x = x^*$. Then one could reach a conclusion that at all places where $(E_F + K_P) \le 0$ (e.g., when $K_P = 0$), the particle could not be observed, since $p$ would be imaginary. These locations are physically forbidden. An example for this status is the micro-harmonic oscillator for which $K_P$ approaches zero at boundaries, but the energy of field $E_F$ is negative. $\chi^2$ is also equal to zero at boundaries. (See section 3.2 below.)

2) $K_P|_{x=x^*} \ne 0$, but $\chi'^2|_{x=x^*}$ is minimum (equal to zero); so, the value of $K_{PF}$ is reduced. The probability of finding the particle in such places is great. Yet, the second condition happens when $\chi(\ne 0)$ has a maximum or minimum at $x = x^*$. Hence, $\chi^2$ is always maximum.

3) $K_P|_{x=x^*} \ne 0$ and $\chi'^2|_{x=x^*}$ is maximum; so, the value of $K_{PF}$ increases. The probability of finding the particle at such places is small. This situation happens when $\chi^2$ is minimum (equal to zero), for example at nodes. For, when $\chi'^2$ is maximum, $\frac{d\chi'^2}{dx} = 2\chi'\chi'' = 0$, which leads to the conclusion that $\chi'' = 0$ (if $\chi' \ne 0$). This in turn entails that $\chi$ should be zero (according to (7)). Hence, $\chi^2$ is minimum.

The above conditions show a very subtle and significant harmony between the values of $K_{PF}$ and $\chi^2$: when $K_{PF}$ is reduced, $\chi^2$ is maximum and vice versa. Even, in some problems when $K_{PF}$ goes to zero, the value of $\chi^2$ changes in accordance with the physical implications of problem [2]. Generally speaking, one can conclude that $|\chi|^2$ is of type of a *probability field* (with dimension of (length)$^2$) and, accordingly, $|\psi|^2$ is a probability *density* (with dimension of (length)$^{-1}$). Correspondingly, the coefficient $\mathcal{N}$ in (8) is a normalization factor. The same interpretation is true for $|X|^2$ and $|\Psi|^2$, because each time-dependent $X(x,t)$ (or $\Psi(x,t)$) can be expanded in terms of stationary fields (or stationary wave functions).

At last, it is substantial to notice that one can never see a PF system in its integrated form. A PF system is in fact a particle enfolded by an energetic field which prevents us to observe the particle without intervention [1]. Of course, we can detect a particle and observe its location in space. In this way, however, we unfold a particle aspect of the hidden PF system which has all known particle properties without any energetic field associated with it.

## 2.2 Time-dependent Schrödinger equation

If $|X|^2$ is a probability field with an objective interpretation, it is legitimate to assume that, the same as the flow of a physical fluid, there should be a *probability flux* obeying the continuity equation.[4] Corresponding to $|X|^2$, one can define a probability density $\rho = |\Psi(x,t)|^2$ where $\Psi(x,t)$ is a time-dependent wave function. Then, alternatively, we have

$$\frac{\partial \rho}{\partial t} + \frac{\partial j}{\partial x} = 0 \tag{19}$$

In above relation, $j(x,t)$ is the probability flux satisfying the following relation, just as in fluid dynamics:

$$\int_{\Omega(x)} dx\, j(x,t) = \frac{\langle p \rangle}{m} \tag{20}$$

where $\langle p \rangle$ is the average value of $p$ (de Broglie momentum) appeared in (6). The integral in (20) is only over the position space $\Omega(x)$. Regarding the assumption that $|\psi|^2$ is a probability density, for a particle subjected to a conservative force, one can obtain $\langle p \rangle$ using time-independent Schrödinger equation $\psi'' = -\frac{p^2}{\hbar^2}\psi$. (See appendix A.) As a result, having the definition of a momentum operator for a one-dimensional system as $\hat{p} := (-i\hbar \frac{\partial}{\partial x})$ (relation (A-3)), for any general state $\Psi(x,t)$, $\langle p \rangle$ is equal to:

$$\langle p \rangle = -i\hbar \int_{\Omega(x)} dx\, \Psi^*(x,t) \frac{\partial}{\partial x} \Psi(x,t) \tag{21}$$

Subsequently, one can show that:

$$\begin{aligned}\frac{\langle p \rangle}{m} &= \frac{1}{2m}(\langle p \rangle + \langle p \rangle^*) \\ &= \frac{-i\hbar}{2m} \int_{\Omega(x)} dx\, \left[\Psi^*(x,t) \frac{\partial}{\partial x} \Psi(x,t) - \Psi(x,t) \frac{\partial}{\partial x} \Psi^*(x,t)\right]\end{aligned} \tag{22}$$

Comparing with (20), we deduce that

$$j(x,t) = \frac{-i\hbar}{2m}\left[\Psi^*(x,t) \frac{\partial}{\partial x} \Psi(x,t) - \Psi(x,t) \frac{\partial}{\partial x} \Psi^*(x,t)\right] \tag{23}$$

Now, from (19), one can write:

$$\begin{aligned}\frac{\partial \rho}{\partial t} &= \Psi^*(x,t) \frac{\partial}{\partial t} \Psi(x,t) - \Psi(x,t) \frac{\partial}{\partial t} \Psi^*(x,t) \\ &= \frac{i\hbar}{2m}\left[\Psi^*(x,t) \frac{\partial^2}{\partial x^2} \Psi(x,t) - \Psi(x,t) \frac{\partial^2}{\partial x^2} \Psi^*(x,t)\right]\end{aligned} \tag{24}$$

If we consider each term in first relation of (24) to be equal with its corresponding term in the second one and multiplying both sides by $(-i\hbar)$, one can show that:

---

[4] An objective probability interpretation, here, means that there exist real, observer-independent properties attached to the very concept of probability, beyond the mere subjective knowledge one can gain from a stochastic examination. In our approach, the probability fields are another facet of an unified reality in which objective physical notions like matter, energy and motion are involved. So, while we might be *ignorant* of some of these properties (e.g., the energy of field) at a given level of examination, their real existence would have effectively outward appearances somewhere in practice (see, e.g., section 4).

$$i\hbar \frac{\partial}{\partial t} \Psi(x,t) = \frac{-\hbar^2}{2m} \frac{\partial^2}{\partial x^2} \Psi(x,t) + V_P \Psi(x,t) \tag{25}$$

This is the time-dependent Schrödinger equation. The $X$-functions satisfy the same equation. The second term in r.h.s of (25) has been introduced according to the requirement that for stationary states where $\Psi(x,t) = \psi(x)f(t)$, the time-dependent Schrödinger equation should be converted to the time-independent equation (25).

In quantum mechanics. when a *system* is described by the eigenfunctions of a linear Hermitian operator corresponding to a physical observable, the observable is a *discernible* property in practice. The notion of system, however, is not clearly depicted in quantum mechanics. Knowing the nature of a micro-system clearly in our approach, we can now explain why, e.g., the energy of the PF system $E$ (and no other quantity) is a discernible quantity in practice for the stationary states. The total energy $E$ is practically a discernible property, because its mean value $\langle E \rangle = \langle V_P \rangle + \langle \frac{p^2}{2m} \rangle$ is the same value defined for the energy of the whole PF system in (15) for stationary fields. This is a general rule. Whenever the mean value of a quantity in a given eigenstate is the same value obtained for the PF system, one can assert that it is a discernible property in practice. So, e.g., the de Broglie momentum $p$ is not a discernible quantity in practice (while it is an observable), except for the stationary fields where $\langle p \rangle = p$, that is for a free PF system (see [1]).

## 2.3 The $N$-particle $3N$-dimensional system

Let us first consider a system containing only two particles each in a three-dimensional Euclidean space. The associated field of particles is represented by

$$X = X(\vec{r}_1(t), \vec{r}_2(t), t) \tag{26}$$

where $\vec{r}_i(t)$ is the local vector of the $i$th particle ($i = 1, 2$). Then, we expect the stationary fields $\chi$ satisfy the time-independent Schrödinger equation:

$$\left[ -\sum_{i=1}^{2} \frac{\hbar^2}{2m_i} \nabla_i^2 + V_P(\vec{r}_1, \vec{r}_2) \right] \chi = E\chi \tag{27}$$

where $(E - V_P) = \sum_i \frac{p_i^2}{2m_i}$ and $\vec{p}_i$ is the de Broglie momentum vector of the $i$th particle.

For an entangled state where $\chi$ in (27) is not factorized, the associated field encompasses both particles. The spatial entangled states, however, are resulted from the fact that the potential term for particles $V_P(\vec{r}_1, \vec{r}_2)$ is not decomposed to independent terms. Yet, As particles fly apart, the spatial interactions between them vanish and each particle experiences a local force. Thus, for well-separated particles, the associated field also factorizes to local fields and a two-particle problem reduces to two distinct one-particle problems. There is no room for non-local actions here.

Similar to the relation (4), we can define the velocity of the stationary field as $v_F = \left| \frac{d\chi}{dt} \right|$. Nevertheless, for most two-body systems in physics, we can define $V_P$ as a function of relative coordinates only. A relative (internal) vector $\vec{r}$ is defined as $\vec{r} = \vec{r}_2 - \vec{r}_1$ with coordinates $x$, $y$,

and $z$. Subsequently, one can show that $v_F = |\dot{\chi}| = |\vec{v}_\mu \cdot \vec{\nabla}\chi|$, where $v_\mu = |\vec{\dot{r}}| = (v_x^2 + v_y^2 + v_z^2)^{1/2}$, $\mu$ is the reduced mass and $\vec{\nabla}$ is defined in relative coordinates.[5]

The internal force the field is subjected to can be represented in relative coordinates according to the following relation:

$$f_{int,F} = \mu v_\mu^2 \nabla^2 \chi(x,y,z) + \vec{f}_P(x,y,z) \cdot \vec{\nabla}\chi(x,y,z)$$
$$= -\mu \bar{\omega}^2 \chi(x,y,z) + \vec{f}_P(x,y,z) \cdot \vec{\nabla}\chi(x,y,z) \tag{28}$$

where $\bar{\omega} = v_\mu k$, $k = \frac{p}{\hbar}$ and $\nabla^2 \chi = -k^2 \chi$ in relative (internal) coordinates. The first term in (28) contains an oscillating-like force similar to the one-dimensional case.

Resembling the relation (10) but for stationary fields, we can express the internal kinetic energy of the whole PF system as $K_{int,PF} = \frac{1}{2}\mu \dot{q}_{int}^2$, where

$$\dot{q}_{int}^2 = (v_\mu^2 + |\dot{\chi}|^2) \tag{29}$$

In above equation, $|\dot{\chi}| = \left|\sum_{\alpha=x,y,z}(\frac{\partial \chi}{\partial \alpha})v_\alpha\right|$. In spherical polar coordinates, it can be expressed as:

$$|\dot{\chi}| = \left|(\frac{\partial \chi}{\partial r})\dot{r} + (\frac{\partial \chi}{\partial \theta})\dot{\theta} + (\frac{\partial \chi}{\partial \varphi})\dot{\varphi}\right| \tag{30}$$

From (29), it is possible to obtain an expression for the internal trajectories of the PF system:

$$q_{int} = \int dt (v_\mu^2 + |\dot{\chi}|^2)^{1/2} \tag{31}$$

In the hydrogen atom problem, we will obtain a more explicit form for internal trajectories illustrated in (31) (see subsection 3.3).

Now, we extend our formalism to describe a system composed of $N$-particles. The associated field of particles is defined in a $3N$-dimensional configuration space:

$$X = X(\vec{r}_1(t), \vec{r}_2(t), \ldots, \vec{r}_N(t), t) \tag{32}$$

and the stationary fields satisfy the same equation as (27), but for $N$-particles.

Let us define the position of an individual PF system in rectangular Cartesian coordinates as $q^{(j)}(\vec{r}_j(t), t)$, where $j = 1, 2, \ldots, N$ denotes the number of particles. Since $|\Psi(\vec{r}_1(t), \vec{r}_2(t), \ldots, \vec{r}_N(t), t)|^2$ is a probability density (corresponding to the probability field $|X(\vec{r}_1(t), \vec{r}_2(t), \ldots, \vec{r}_N(t), t)|^2$), one can define an associated marginal field $|X^{(j)}(\vec{r}_j(t), t)|^2$ for the $j$th particle as:

---

[5]Here, in general, we analyze the many-body problem in the center-of-mass frame, where the total momentum of particles is assumed to be zero and therefore the kinetic energy of the system only includes the motion of particles relative to the center-of-mass.

$$|X^{(j)}(\vec{r}_j(t),t)|^2 = \left(\frac{A}{\mathcal{N}^{(j)}}\right)^2 |\Psi^{(j)}(\vec{r}_j(t),t)|^2 \qquad (33\text{-a})$$

where $\mathcal{N}^{(j)}$ is the normalization constant of the wave function $\Psi^{(j)}$, and $A$ has been defined in (8), having the dimension of length. For $|\Psi^{(j)}(\vec{r}_j(t),t)|^2$ we have:

$$|\Psi^{(j)}(\vec{r}_j(t),t)|^2 = \int_{R_1} \cdots \int_{R_N} \prod_{k=1;k\neq j}^{N} d\vec{r}_k |\Psi(\vec{r}_1(t),\vec{r}_2(t)\ldots,\vec{r}_j(t),\ldots,\vec{r}_N(t),t)|^2 \qquad (33\text{-b})$$

where the integrals extend over all configuration space except for the coordinates of $\vec{r}_j$.

Correspondingly, we can describe kinetic energy of the $j$th PF system as

$$K_{PF}^{(j)} = \frac{1}{2} m_j \left|\vec{\dot{q}}^{(j)}(\vec{r}_j(t),t)\right|^2 \qquad (34)$$

where $\vec{\dot{q}}^{(j)}(\vec{r}_j(t),t)$ is velocity vector of the $j$th PF system, which its value is equal to:

$$\dot{q}^{(j)}(\vec{r}_j(t),t) = \left|\vec{\dot{q}}^{(j)}(\vec{r}_j(t),t)\right| = \sqrt{\left(v_p^{(j)2} + |\dot{X}^{(j)}(\vec{r}_j(t),t)|^2\right)} \qquad (35)$$

Here, $v_p^{(j)}$ is the velocity of the $j$th particle which can be obtained by solving the classical equations of motion. From (35), one can also obtain the local position of each individual PF system as

$$q^{(j)}(\vec{r}_j(t),t) = \int dt\, \dot{q}^{(j)}(\vec{r}_j(t),t) \qquad (36)$$

For stationary fields, one can define the total energy of system as:

$$E = V_P(\vec{r}_1(t),\vec{r}_2(t),\ldots,\vec{r}_N(t)) + \left(E_F + \sum_{j=1}^{N} \frac{p_P^{(j)2}}{2m_j}\right) = V_P + \sum_{j=1}^{N} \frac{p_j^2}{2m_j} \qquad (37)$$

where $K_P = \sum_{j=1}^{N} \frac{p_P^{(j)2}}{2m_j}$ is the total kinetic energy of $N$-particles and $p_j$ is the de Broglie momentum defined for the $j$th particle. The energy of field $E_F$ is also defined as $E_F = V_F + \sum_{j=1}^{N} \frac{1}{2} m_j |\dot{\chi}^{(j)}(\vec{r}_j(t))|^2$ where $V_F$ is the potential term of the field and

$$K_F = \sum_{j=1}^{N} \frac{1}{2} m_j |\dot{\chi}^{(j)}(\vec{r}_j(t))|^2 \qquad (38)$$

is the kinetic term. Alike the relation (15), the total observable energy $E$ can then be read out as

$$E = E_P + E_F = V_{PF} + K_{PF} = V_{PF} + \sum_{j=1}^{N} \frac{1}{2} m_j [\dot{q}^{(j)}(\vec{r}_j(t))]^2 \qquad (39)$$

where $V_{PF}$ is the potential term of the whole PF system expressible in terms of $E$ and $K_{PF}$, if velocities $\dot{q}^{(j)}(\vec{r}_j(t))$ are known.

One may wonder how $|\chi|^2$ in a *3N-dimensional configuration space* can be regarded as an *objective* probability field endowed with energy $E_F$, while this high-dimensional space has no objective meaning in external world. The reader should note that the probability field $|\chi(\vec{r}_1(t),\vec{r}_2(t),\ldots,\vec{r}_N(t))|^2$ has two facets. On one hand, in a $3N$-dimensional configuration space, it represents a probability density $|\psi(\vec{r}_1(t),\vec{r}_2(t),\ldots,\vec{r}_N(t))|^2$ from which one gains knowledge about possible locations one can find particles. This is very usual in all statistical approaches to define a probability density function in spaces with dimensions greater that 3, because it is only a mathematical function.

On the other hand, the field has an energy $E_F$ which is determined by knowing the complete from of $\chi$. This aspect too is not unusual, since the very notion of potential energy in classical physics is a field of energy defined in a high-dimensional space (see, e.g., $V_P(\vec{r}_1(t),\vec{r}_2(t),\ldots,\vec{r}_N(t))$ in (37)).

This is a main problem in all Bohmian approaches of quantum theory, however [4]. In Bohm's theory, the wave function $\Psi$ is defined in a $3N$-dimensional space (for a $N$-particle system) which either is supposed to have an impact on particles through a so-called quantum force, or at least its phase determines the velocity fields of particles. This is peculiar, since the particles are assumed to be in three-dimensional Euclidean space, but their properties are *determined* or *affected* by a wave which is defined in a high-dimensional configuration space. In our description, however, there is no action or interaction between the particle and its associated field. The full description of system, here, is based on $q$-functions which are defined in three-dimensional Euclidean space for each individual PF system (see relation (36)).

Another important point is that neither $|X^{(j)}(\vec{r}_j(t),t)|^2$, nor $q^{(j)}(\vec{r}_j(t),t)$ are independent fragmentary notions, although they are defined for local fragments. All quantities which are considered for individual parts in our approach, are in fact *contextual* properties depending on the whole behavior of the global system, since they are derived from the whole. For example, for two identical particles (say, two electrons), both having the same local forces, same velocities and so forth, but different environments, the entire probability field is not the same. Hence their complete physical situations delineated by a PF description are not the same. Only when *all* particles are independent of each other, i.e., when they have no local interactions with one another, this peculiar context-dependent feature disappears.

To explain the subject more obviously, let us consider, for example, a system composed of three interacting particles. Suppose that particle 1 only interacts with particle 2, but particle 2 interacts with both other particles. Suppose also that all interactions are local and conservative. In classical mechanics, the physical state of particle 1 is independent of particle 3, but is contingent upon the status of the second particle. Taking this picture from classical mechanics and assuming that the only interactions between the particles are those described by classical local forces, in the formalism depicted here, we see that there should also exist an objective probability field $|\chi(\vec{r}_1(t),\vec{r}_2(t),\vec{r}_3(t))|^2$ associated with *all* particles which satisfies the time-independent Schrödinger equation in which *all* interactions must be taken into account (say, by introducing $V_P = V_{12} + V_{23}$ in (37)). In this example, the field can be factorized to $|\chi_{12}(\vec{r}_1(t),\vec{r}_2(t))|^2|\chi_{23}(\vec{r}_2(t),\vec{r}_3(t))|^2$. Then, the actual physical state of each part is represented by an appropriate $q$-function in (36) which includes the particle-picture as well as its associated field-picture totally. The local marginal field $|\chi^{(1)}(\vec{r}_1(t))|^2$, however, is obtained by integrating over *both* spatial variables $\vec{r}_2$ and $\vec{r}_3$. In effect, it is influenced (albeit, not mechanically) by what happens to particle 3, although there is no interaction between particles 1 and 3. Hence, the PF description is not fragmentary, but it is a holistic delineation. This holistic feature is appeared not because of the existence of non-local interactions (like what, e.g. the Bohm theory posits [4]), but it is the result of the whole influences a local (marginal) field with a real objective character receives from its environment.

As a result, while in classical mechanics, only local interactions determine the physical status of particles and in quantum mechanics only the wave function (which is only a mathematical tool for predicting the probabilities of getting various results in repeated experiments) represents the physical state of a system, here, both theories play role in explaining the micro-phenomena in a coherent and consistent way.

This is indeed the local contextual feature of our theory which makes it to be distinguished from both classical mechanics and quantum theory. The notion of context-dependency, however, may enter from a different sight too, when the measuring contexts are mutually exclusive in practice. For example, one cannot simultaneously measure the spin components of a particle in a single experiment. Thus, corresponding to different spin components, different sets of (hidden) spin variables may be used in an underlying theory to reproduce the experimental results. This distinct feature makes a fundamental theory not be captured by the so-called no-go theorems such as the Kochen-Specker [5] and Bell theorems [6]. Based on the assumptions of locality and contextuality, we have recently suggested a spin model which could reproduce Bell correlations based on the principle of causality [7].[6]

## 3  Stationary fields

### 3.1  Particle in a one-dimensional box

Let us consider a particle confined to a one-dimensional box along the $x$-direction within $0 \leq x \leq a$ where the particle subjected to no force. At boundaries, the potential energy $V_P$ equals infinity. According to (7), we have

$$\chi_n(x(t)) = A_n \sin\left(\frac{n\pi x(t)}{a}\right) \tag{40}$$

where $A_n$ is the amplitude of the field $\chi_n$ (defined also in (8)), and $n = 1,2,\ldots$ is the energy quantum number. The energy of the system is also obtained as $E_n = \frac{p_n^2}{2m} = \frac{n^2 h^2}{8ma^2}$ where $p_n$ is the de Broglie momentum in (6). Considering $x(t) = v_P t + x_0$, where $x_0$ is the initial position of particle, one can also write (40) as:

$$\chi_n(t) = A_n \sin(\bar{\omega}_n t + \phi) \tag{41}$$

where $\bar{\omega}_n = k_n v_P$, $k_n = \frac{n\pi}{a}$ and $\phi = k_n x_0$. The relation (41) is the answer of (9) too, when one puts $f_P = 0$ in it. Taking into account the above relation as the answer of field's equation of

---

[6]The original EPR proposal, however, is a special case in which the above mentioned context-dependency plays no role. For more explanation, see [1].

motion (9), one can simply show that the energy of field is equal to:

$$E_{n,F} = \frac{1}{2} m \bar{\omega}_n^2 A_n^2 \tag{42}$$

Using the relation (4), the kinetic energy of field can be introduced as:

$$K_{n,F} = \frac{1}{2} m \bar{\omega}_n^2 A_n^2 \cos^2\left(\frac{n\pi x(t)}{a}\right) \tag{43}$$

And the potential of field is:

$$V_{n,F} = \frac{1}{2} m \bar{\omega}_n^2 A_n^2 \sin^2\left(\frac{n\pi x(t)}{a}\right) \tag{44}$$

Now, considering the equivalence relation $E_n = E_P + E_{n,F}$, where $E_P = \frac{p_P^2}{2m}$, one can obtain

$$E_n = \frac{E_P}{1 - \frac{p_P^2 A_n^2}{\hbar^2}} \tag{45}$$

which should be always positive. So, we have

$$p_P^2 A_n^2 \leq \hbar^2 \tag{46}$$

The inequality (46) shows that for a particle with high velocity, the magnitude of the field's amplitude is reduced. As the kinetic energy of the particle grows, the particle's energy $E_P$ approaches the observable energy of the system $E_n$ and the energy of field will vanish.

From (45), one can show that

$$A_n = \pm \frac{\hbar}{p_P}\left(1 - \frac{p_P^2}{p_n^2}\right)^{1/2} \tag{47}$$

where $p_n^2 = \frac{n^2 \hbar^2}{4a^2}$. One can also check that we should always have

$$p_P^2 \leq p_n^2 \tag{48}$$

It is now apparent that when $p_P^2$ approaches $p_n^2$ (or, equivalently, when $E_P$ approaches $E_n$), then $A_n \to 0$ and $E_F$ becomes zero and we will have a field without energy characterized as $\psi(x)$ only. This is the classical limit introduced before in section 2. A physical instance of this situation is when $n \to \infty$. When we give energy to system, the value of $E_P$ and the value of $E_n$ both will increase. According to (46), when $p_P^2$ increases, the value of $A_n^2$ should inevitably decreases. Then, $E_{n,F}$ can be neglected, and $E_n; E_P$, so that, $A_n \to 0$ in (47). In effect, we will have a particle with a classical uniform spatial distribution $\rho_{n \to \infty} = \frac{1}{a}$.

Now, let us obtain the trajectories of the whole PF system according to (12), where $\chi = \chi_n(x(t))$ in (40). Here, one can show that

$$q_n(x) = g_{n,PF} \int dx (1 + \chi_n'^2(x))^{\frac{1}{2}} = g_{n,PF} \int dx \left[\sin^2\left(\frac{n\pi x}{a}\right) + \frac{p_n^2}{p_P^2}\cos^2\left(\frac{n\pi x}{a}\right)\right]^{\frac{1}{2}} \qquad (49)$$

This is an elliptic integral which we prefer to solve it with an appropriate physical assumption. The main assumption is that we can expand $f_n(x) = (1 + \chi_n'^2(x))^{1/2}$ about the small values of $\chi_n'$ corresponding to the small values of $A_n$. From a physical point of view, this assumption means that the energy of field is not so considerable comparing to the energy of particle. This assumption is justified, because the PF system has a particle character indeed. This is the particle which shares some of its energy with its surrounding field. Yet, the energy contribution of the field is not so significant. Otherwise, the particle nature of the PF system becomes problematic. Thus, we can use the following approximation

$$f_n(x) = (1 + \chi_n'^2(x))^{1/2} \simeq 1 + \frac{\chi_n'^2(x)}{2} \qquad (50)$$

from which it is straightforward to show that

$$q_n(x) \simeq g_{n,PF}\left[x + \frac{1}{2}\left(\frac{p_n^2}{p_P^2} - 1\right)\int dx\cos^2\left(\frac{n\pi x}{a}\right)\right] \qquad (51)$$

So, defining $g_{n,PF} = \frac{4p_P^2}{p_n^2 + 3p_P^2}$, we get

$$q_n(x) \simeq x + \frac{a}{2n\pi}\left(\frac{p_n^2 - p_P^2}{p_n^2 + 3p_P^2}\right)\sin\left(\frac{2n\pi x}{a}\right) \qquad (52)$$

which demonstrates the paths of the PF system in a one-dimensional box with width $a$. The justification of approximation (50) for reaching the trajectories (52) has been elaborated in Appendix B. One should also notice that as $p_P^2 \to p_n^2$ (in the classical limit), $g_{n,PF} \to 1$ and $q_n(x) \to x$.

In figures 1-a, b and c, $q_n(x)$ is drawn for $n = 1, 2,$ and $3$, respectively, in terms of $x$. One can draw the trajectories in terms of $t$ too, considering $x(t) = v_P t + x_0$, where $v_P$ and $x_0$ should be chosen arbitrarily. It is significant to note that while in the classical view, the trajectory of the particle is just a straight line, here the trajectories are curved. The curvature of space for the PF system is due to addition of a new direction caused by the presence of a transverse field denoted by $\chi_n(x)$ (see also relation (11)).

The wave function of the system is introduced as

$$\psi_n(x) = \frac{\mathcal{N}}{A_n}\chi_n(x) = \mathcal{N}\sin\left(\frac{n\pi x}{a}\right) \qquad (53)$$

where $\int_0^a dx\,|\psi_n(x)|^2 = 1$ and so $\mathcal{N} = \sqrt{\frac{2}{a}}$. In figures 1 to 3 it is obvious that in each point where $|\psi_n(x)|^2$ has a maximum or minimum, $q_n(x)$ has an inflection point. Rewriting the relation (14) for the case of translational motion with $f_P = 0$, we obtain:

$$f_{PF} = m\ddot{q}_n(x) = g_{n,PF}\left[\frac{mv_P^2 \chi_n' \chi_n''}{(1+\chi_n'^2)^{1/2}}\right] \tag{54}$$

At nodes, where $\chi_n = 0$ and so $\chi_n'' = 0$, according to (54), $\ddot{q}_n(x) = 0$ and we have an inflection point for $q_n(x)$. Also, the velocity of the PF system is equal to

$$\dot{q}_n(x) = g_{n,PF}\ v_P\sqrt{1 + \chi_n'^2(x)} \tag{55}$$

which shows that at nodes where $\chi_n'^2(x)$ is maximum, $\dot{q}_n(x)$ and the kinetic energy of field reach their maximum value. Therefore, the probability of finding the particle in the vicinity of nodes is very small, because the maximum value of the kinetic energy of field (i.e., $K_F = K_p|\chi_n'|^2$) prevents the particle to stay there. In other words, the velocity of the PF system is so high that lessens the chance of observing the particle in such points, when one measures particles position.

When $|\psi_n(x)|^2$ has a maximum, $\psi_n'(x) = 0$ and hence $\chi_n'(x) = 0$. Consequently, both the kinetic energy of field and $\dot{q}_n(x)$ reach their minimum values and $\ddot{q}_n(x) = 0$ in (54) indicating the existence of another inflection point in the $q$-curves. In such points, the kinetic energy of field decreases and there will be a more chance the particle can be found there.

### 3.2 The two-particle three-dimensional isotropic harmonic oscillator

We consider the three-dimensional isotropic harmonic oscillator for which the classical potential energy is defined as:

$$V_\mu = \frac{1}{2}\mu\omega_0^2(r - r_{eq})^2 \tag{56}$$

where $\mu$ is the reduced mass, $r = |\vec{r}_2 - \vec{r}_1|$ and $r_{eq}$ are internal and equilibrium distances of two particles, respectively and $\omega_0$ is the spring frequency which is supposed to have the same magnitude in all directions. Defining $\bar{r} = r - r_{eq}$, the classical internal dynamics of the two-particle system (hereafter, we call it $\mu$-particle) can be written as

$$\mu\frac{d^2\bar{r}}{dt^2} = f_\mu(\bar{r}) \tag{57}$$

where $f_\mu(\bar{r}) = -\mu\omega_0^2\bar{r}$.

We also introduce the associated internal fields of the oscillator as

$$\chi_{n,l,m_l}^{osc}(\bar{r},\theta,\varphi) = \chi_n(\bar{r})Y_{l,m_l}(\theta,\varphi) \tag{58}$$

which are the solutions of the time-independent Schrödinger equation. In (58), $\chi_n(\bar{r})$ is a radial field which depends on the energy quantum number $n$ ($n = 0,1,2,...$), and $Y_{l,m}(\theta,\varphi)$ denotes the well-known spherical harmonic functions where $l$ and $m_l$ are angular momentum and its z-component quantum numbers, respectively.

From the equation (57), one can find the following solutions for the internal distance and momentum variables $\bar{r}(t)$ and $p_\mu(t)$, respectively:

$$\bar{r}(t) = L\cos(\omega_0 t + \vartheta);$$
$$p_\mu(t) = \mu \frac{d\bar{r}}{dt} = -\mu\omega_0 \frac{d\bar{r}}{dt}\sin(\omega_0 t + \vartheta) = -\mu\omega_0\sqrt{L^2 - \bar{r}^2(t)} \tag{59}$$

where $L$ is the amplitude of oscillation and $\vartheta$ is a phase constant.

The energy of field can be represented as

$$E_{n,F} = E_n - E_\mu = \frac{\hbar\omega_0}{2}[2n + 1 - \alpha L^2] \tag{60}$$

where $E_n = \hbar\omega_0(n + \frac{1}{2})$ is the energy of the whole oscillating PF system, $\alpha = \frac{\mu\omega_0}{\hbar}$ and $E_\mu = \frac{1}{2}\mu\omega_0^2 L^2$ is the internal energy of $\mu$-particle. Here, we are assuming that $L^2 \geq \frac{2n+1}{\alpha}$. Hence, there is no classically forbidden region where the PF system cannot go there, because $E_n \leq E_P$. The energy of field is negative and varying depending on various $n$ states, but the energy of $\mu$-particle is constant. To prepare an excited PF system, we must give energy to the whole system. As the system takes energy, the energy of field $E_{n,F}$ approaches zero and at the same time, the value of $E_n$ positively increases and becomes near to $E_\mu$.

Let us suppose that the energy of field would be zero for $n \geq n_{\max}$. Then, we have

$$L^2 = \frac{2n+1}{\alpha}; \quad for\ n \geq n_{\max} \tag{61}$$

For a hydrogen molecule, e.g., with $\alpha \approx 10^{20}$, assuming that $n_{\max} \approx 50$, one gets $L \simeq \pm 1$ nm.

Beyond the $n_{\max}$, if we still give energy to the system, both $E_n$ and $E_\mu$ will increase equivalently, but the energy of field remains zero. In the classical limit $E_{n,F} \to 0$, we have only a particle entity which is statistically described by a classical probability distribution.

The stationary probability fields go to zero, when particle reaches its boundaries. It is because all radial distributions $\chi_n^2(\bar{r}(t))$ are proportional to a factor of $\exp(-\alpha\bar{r}^2(t))$, where at $\bar{r} \to L$ comes near to zero. For example, for a typical value of $\alpha L^2 \approx 101$ (see relation (61) for $n_{\max} \approx 50$), $\chi_n^2(L) \propto \exp(-\alpha L^2) \approx \exp(-101); 10^{-44}$ which is physically negligible. The same situation holds true for the kinetic energy of field. Regarding the relation (30), one can define:

$$K_{n,F} = \frac{1}{2}\mu|\dot{\chi}_{n,l,m_l}^{osc}|^2 = \frac{1}{2}\mu\left(\frac{d\bar{r}}{dt}\right)^2\left(\frac{d\chi_n(\bar{r})}{d\bar{r}}\right)^2 |Y_{l,m_l}(\theta,\varphi)|^2$$
$$= K_\mu \chi_n'^2(\bar{r})^2|Y_{l,m_l}(\theta,\varphi)|^2 \tag{62}$$

where, as is obvious from the classical Newton's equation (57), $\dot{\theta}$ and $\dot{\varphi}$ are supposed to be zero. One can see that $K_{n,F} \to 0$, as $\bar{r} \to L$.

On the other hand, $p_\mu(t)$ comes to zero, when $\bar{r} = L$. Yet, at boundaries the potential of field reaches its minimum value $E_{n,F} < 0$ and $p^2 = 2\mu E_{n,F}$, where $p$ is the de Broglie momentum. Since $p$ is physically significant in our approach, it is not possible it takes imaginary values. Correspondingly, when $K_\mu = 0$ (except for the cases in which $E_{n,F} = 0$, i.e., in the classical limit $n \to n_{max}$), the probability densities (i.e., $|\psi_{n,l,m_l}^{osc}|^2 \propto |\chi_{n,l,m_l}^{osc}|^2$) will be zero at boundaries for $n = n_{max}$.

To find the trajectory equations of the oscillating PF system, we first rewrite (29) as

$$\dot{q}_{n,l,m_l}^2(\bar{r}, \theta, \varphi) = g_{PF}^2 \, v_\mu^2 \left(1 + \chi_n'^2(\bar{r})^2 |Y_{l,m_l}(\theta, \varphi)|^2\right) \tag{63}$$

where $v_\mu^2 = \left(\frac{d\bar{r}}{dt}\right)^2$ and we set $g_{PF} = (2\sqrt{\pi})^{-1}$. Integrating over $\theta$ and $\varphi$ in the spherical coordinates and assuming that spherical harmonics $Y_{l,m_l}(\theta, \varphi)$ are normalized, one eventually obtains:

$$q_n^{osc}(\bar{r}) = \int d\bar{r} \left(1 + \frac{\chi_n'^2(\bar{r})}{4\pi}\right)^{1/2} \simeq \int d\bar{r} \left(1 + \frac{\chi_n'^2(\bar{r})}{8\pi}\right) \tag{64}$$

in which we have used the same approximation as (50).

Now, let us calculate the trajectories for the first two states of an oscillating PF system. Calculations for other stationary fields can be made in a similar way. For $n = 0$ and 1, we have, respectively, $\chi_0(\bar{r}) = A_0^{osc} \exp(-\alpha \bar{r}^2/2)$ and $\chi_1(\bar{r}) = A_1^{osc} \bar{r} \exp(-\alpha \bar{r}^2/2)$ where $A_n^{osc}$ is the amplitude of the stationary field for the $n$th state. The trajectory functions for $n = 0$ and 1, can then be calculated as:

$$q_0^{osc}(\bar{r}) \simeq \bar{r} + \frac{(\alpha A_0^{osc})^2 \bar{r}^3}{48\pi} e^{-\alpha \bar{r}^2} + \ldots \tag{65}$$

and

$$q_1^{osc}(\bar{r}) \simeq \bar{r} + \frac{\alpha (A_1^{osc})^2 \bar{r}}{16\pi} e^{-\alpha \bar{r}^2} + \frac{\alpha^3 (A_1^{osc})^2 \bar{r}^5}{80\pi} e^{-\alpha \bar{r}^2} + \ldots \tag{66}$$

respectively. The details of the derivation of above relations are given in Appendix C.

For now, it is useful to estimate the values of $A_0^{osc}$ and $A_1^{osc}$ for a specific system like the hydrogen molecule. Both the $q$-functions in (65) and (66) are zero at $\bar{r} = 0$ (i.e., $r = r_{eq}$) and it is expected they satisfy the boundary condition $q_n^{osc}(\bar{r} = L) = L$. In the first excited state, $q_1^{osc}(\bar{r})$ should also have a near value to $\pm\frac{1}{\sqrt{\alpha}}$ at $\bar{r} = \pm\frac{1}{\sqrt{\alpha}}$, since $\chi_1^2(\bar{r})$ has two similar maximums at $+\frac{1}{\sqrt{\alpha}}$ and $-\frac{1}{\sqrt{\alpha}}$. So, for a hydrogen molecule with $\alpha \approx 10^{20}$ and $L \simeq \pm 1$ nm, one concludes from the relation (66) that

$$q_1^{osc}(\bar{r} = L) \simeq L(1 + 4 \times 10^{-23}(A_1^{osc})^2 + \ldots) \tag{67}$$

and

$$q_1^{osc}\left(\bar{r} = \pm\frac{1}{\sqrt{\alpha}}\right) \simeq \pm\frac{1}{\sqrt{\alpha}}(1 + 8.8 \times 10^{17}(A_1^{osc})^2 + \ldots) \tag{68}$$

Estimating $A_1^{osc} \sim \pm 10^{-10}$ m, we obtain $q_1^{osc}(\bar{r} = L) \simeq L$ in (67), and $q_1^{osc}(\bar{r} = \pm\frac{1}{\sqrt{\alpha}}) \simeq \pm\frac{1.01}{\sqrt{\alpha}}$ in (68). Similarly, for the ground state we have:

$$q_0^{osc}(\bar{r} = L) \simeq L(1 + 6.7 \times 10^{-25}(A_0^{osc})^2 + \ldots) \qquad (69)$$

Comparing this equation with (67), one can appropriately estimate $A_0^{osc} \sim \pm 10^{-9}$ m. Having the same requirements as (67) and (68) for each $q_n^{osc}$, one can show that as the quantum number $n$ increases to higher values ($n \to n_{max}$), $A_n^{osc}$ becomes smaller and smaller. Hence, for $n \to n_{max}$, $q_{n \to n_{max}}^{osc}(\bar{r}) \to \bar{r}$, and we will have only an oscillating particle with a classical distribution.

According to relation (63), the radial kinatic energy of the PF system can be introduced as:

$$K_{n,PF} = K_\mu \left(1 + \frac{\chi_n'^2(\bar{r})}{4\pi}\right) \qquad (70)$$

At nodes where $\chi_n'^2(\bar{r})$ is maximum, the value of $K_{n,PF}$ increases and at places where $\chi_n^2(\bar{r})$ is maximum, it decreases, corresponding to the least and the highest probabilities of finding the particle in such regions.

In effect, the velocity of the oscillating PF system can be approximated to

$$\dot{q}_n^{osc}(\bar{r}) \simeq v_\mu \left(1 + \frac{\chi_n'^2(\bar{r})}{8\pi}\right) \qquad (71)$$

From (71) it is clear that for the first two states, corresponding to the minimum and maximums of the probability fields $\chi_0^2(\bar{r})$ and $\chi_1^2(\bar{r})$, there are one and three inflection points in trajectories $q_0^{osc}(\bar{r})$ and $q_1^{osc}(\bar{r})$, respectively, with the same interpretation depicted before for the particle in box. Yet, to plot the aforementioned trajectories, one should have the complete forms of $q$-functions in (65) and (66). This is a mathematical problem which is not our main concern here.

### 3.3 The hydrogen-like atom

A hydrogen-like atom is composed of an electron of charge $-e$ and a nucleus of charge $+Ze$, where $e = 1.6 \times 10^{-19}$ C. The coordinates of electron relative to the nucleus are denoted as $(x, y, z)$ which are the components of an internal (relative) vector $\vec{r}$. The force exerted on the electron is the Coulomb's law force which for a hydrogen-like atom can be expressed as

$$f = \frac{-Ze'^2}{r^2} \frac{\vec{r}}{r} \qquad (72)$$

where $e' = \frac{e}{\sqrt{4\pi\varepsilon_0}}$ is the proton charge in statcoulombs. The Coulomb force is central for which the potential energy can be obtained as $V = \frac{-Ze'^2}{r}$. Now, we are interested to discuss about a $\mu$-particle moving in a central force field defined in (72). Here, $\mu = \frac{m_e m_N}{m_e + m_N}$ where $m_e$ and $m_N$ are the electronic and nuclear masses, respectively. In this way, for all physical purposes, one can imagine a $\mu$-particle as an electron, because $m_e \cong m_N$ and $\mu; m_e$.

Due to the spherical symmetry which is the typical feature of a central force problem, the angular-momentum of the $\mu$-particle is conserved. Then, the classical internal energy of the hydrogen-like atom can be given as:

$$E_\mu = \frac{1}{2}\mu\dot{r}^2 + \frac{L_C^2}{2\mu r^2} - \frac{Ze'^2}{r} \tag{73}$$

where $L_C^2$ is the square of the angular-momentum magnitude in the classical approach. In a polar coordinates, one can show that

$$L_C = \mu r^2 \dot{\theta} = \mu r v \tag{74}$$

where $\dot{\theta}$ and $v = r\dot{\theta}$ are the angular and tangential velocities, respectively. If we assume that $E_\mu$ is equal to the minimum point of the last two terms $\frac{L_C^2}{2\mu r^2} - \frac{Ze'^2}{r}$, the electron would have a orbital motion with a fixed separation $r$. So, $\dot{r} = 0$ and the internal energy of the $\mu$-particle reduces to:

$$E_\mu = \frac{L_C^2}{2\mu r^2} - \frac{Ze'^2}{r} \tag{75}$$

where $\frac{L_C^2}{2\mu r^2}$ describes the rotational energy of the $\mu$-system with a moment of inertia $\mu r^2$.

From a classical point of view, however, the mechanical stability of the $\mu$-particle orbit requires that the Coulomb force be balanced by the centrifugal force $\mu r \dot{\theta}^2$ due to the circular motion. From this, one concludes that

$$r = \frac{Ze'^2}{\mu v^2} \tag{76}$$

and

$$E_\mu = -\frac{1}{2}\left(\frac{Ze'^2}{r}\right) = -\frac{1}{2}\mu v^2; -\frac{1}{2}m_e\left(\frac{Ze'^2}{\hbar}\right)^2\left(\frac{a_0/Z}{r}\right) \tag{77}$$

where $a_0 = \frac{\hbar^2}{\mu e'^2}$.[7]

The classical dynamics of the $\mu$-particle can be decomposed to three relations corresponding to the three components of $r$. For the $x$-component of $r$, e.g., one can write

$$\mu\frac{d^2x}{dt^2} = f_x = \frac{-Ze'^2}{r^2}\frac{x}{r} = -\dot{\theta}^2 x \tag{78}$$

where $\dot{\theta}^2 = \frac{Ze'^2}{\mu r^3}$ is constant. In accordance with the definition of the $x$-variable in the spherical polar coordinates, one can write the solution of (78) as

$$x = r\sin\theta\cos\varphi \tag{79}$$

where $\theta = \dot{\theta}t + \theta_0$ and $r$ and $\varphi$ are constant. Then, we can obtain the $x$-component of the

---

[7]More accurately, the Bohr radius is defined as $a_0 = \frac{\hbar^2}{m_e e'^2} = 0.52918$ Å. Considering $\mu \simeq m_e$, the two definitions coincide.

tangential velocity $v$ as:

$$v_x = v\cos\theta\cos\varphi \qquad (80)$$

in which $v = r\dot{\theta}$. The other components of $r$ and $v$ can be similarly obtained afterwards.

To know the exact value of $r$, however, we should know the hidden parameter $\dot{\theta}$ in (78). Yet, e.g., for two $\mu$-particles (say, two electrons) with the same quantum energy $E_n$, the angular velocities might have different amounts, so their orbital radii $r$ would be different too. This, in turn, explains why we can prepare hydrogen atoms all with the same total electronic energy $E_n$ in practice, while for each atom the electron may be found in different locations around the nucleus.

Substantially, one should remember that each $\mu$-particle is also associated with an internal field collectively denoted by

$$\chi^{HA}_{n,l,m_l}(r,\theta,\varphi) = \chi_{n,l}(r) Y_{l,m_l}(\theta,\varphi) \qquad (81)$$

where superscript $HA$ is an abbreviation for the hydrogen-atom system, $\chi_{n,l}(r)$ is the radial field of the system and $Y_{l,m_l}(\theta,\varphi) = S_{l,m_l}(\theta) T_{m_l}(\varphi)$ is the spherical harmonics. $T_{m_l}(\varphi) = \frac{1}{\sqrt{2\pi}} \exp(im_l\varphi)$ also denote the eigenfunctions for the $z$-component of angular momentum and $S$ is the theta factor of $Y$. The quantum numbers $n$, $l$ and $m_l$ are defined as in (58), but $n = 1, 2, 3, \ldots$ here. All forms of internal fields $\chi^{HA}_{n,l,m_l}(r,\theta,\varphi)$ are solutions of the time-independent Schrödinger equation written for a hydrogen-like system with the central force (72). Subsequently, the energy of field can be introduced as

$$E_{n,F} = -\frac{1}{2}\mu \left(\frac{Ze'^2}{\hbar}\right)^2 \left[\frac{1}{n^2} - \frac{1}{r}\left(\frac{a_0}{Z}\right)\right] \simeq -\frac{1}{2}m_e \left(\frac{Ze'^2}{\hbar}\right)^2 \left[\frac{1}{n^2} - \frac{1}{r}\left(\frac{a_0}{Z}\right)\right] \qquad (82)$$

where $E_{n,F} = E_n - E_\mu$ and $E_n = -\frac{1}{2}\mu\left(\frac{Ze'^2}{n\hbar}\right)^2$ is the quantum energy of the PF system (i.e., the electron with its associated field (81)). Accordingly, the following situations can be recognized for a given energy level $n$:

- The electron may be found at locations where $r < \frac{n^2 a_0}{Z}$. In this case, the energy of the electron has been more negative compared to the total energy $E_n$. Consequently, the energy of its associated field has been positive.
- The electron may be found at locations where $r \geq \frac{n^2 a_0}{Z}$. Then, one can infer that $E_{n,F} \leq 0$ and $E_\mu \geq E_n$.

The whole PF system can be viewed as an isolated atom with a definite amount of energy $E_n$ in stationary states. The balance of energy between the particle and its associated stationary field is a consequence of the conservation of energy of the whole PF system which in turn is resulted from the fact that when the particle is subjected to a conservative force (like (72)), the total energy of system should be also conserved.[8] In effect, the *active* role belongs to the particle itself and this is the physical status of particle that determines the physical behavior of its allied field.

Thus, a more (less) negative energy of particle implies that electron will be definitely observed nearer (farther apart) the nucleus. For an electron far apart the nucleus (i.e., for large values of $r$), the energy of electron has been more shared with its surrounding, so $E_{n,F}$ is nearly the same as $E_n$. For small values of $n$, however, this situation is not likely to happen. Since, generally speaking, the magnitude of the energy of field has been mostly assumed to be not so large compared to the particle's energy. For higher energy levels, on the other hand, the magnitudes of $E_{n,F}$ and $E_n$ are both negligible, so the electron has a more chance to be found at distant locations around the nucleus. This kind of anticipation is in full agreement with the predictions of quantum mechanics.

Yet, the energy of electron cannot be determined *a priori*, just as the value of the angular velocity $\dot{\theta}$ could not be assigned in advance. So, any prediction about the location of electron is made by using the probability densities $|\psi_{n,l,m_l}^{HA}(r,\theta,\varphi)|^2$ which are proportional to the probability fields $|\chi_{n,l,m_l}^{HA}(r,\theta,\varphi)|^2$. On average, the energy of field is always zero, because in (82) we have $\langle \frac{1}{r} \rangle_{n,l,m_l} = \frac{Z}{a_0 n^2}$. Thus

$$\langle E_\mu \rangle_{n,l,m_l} = E_n \tag{83}$$

Using the relations (30) and (81), one can obtain the kinetic energy of field as

$$K_{n,F} = \frac{1}{2}\mu |\dot{\chi}_{n,l,m_l}^{HA}(r,\theta,\varphi)|^2 = \frac{1}{4\pi}\mu \chi_{n,l}^2(r) S_{l,m_l}^2(\theta) \dot{\theta}^2 \tag{84}$$

Hence, the kinetic energy of the PF system can be introduced as

$$K_{n,PF} = \frac{1}{2}\mu r^2 \dot{\theta}^2 \left(1 + \frac{1}{2\pi r^2}\chi_{n,l}^2(r) S_{l,m_l}^2(\theta)\right) \tag{85}$$

Also, we have

$$\chi_{n,l}(r) = \frac{A_n^{HA}}{\mathcal{N}_n} R_{n,l}(r) \tag{86}$$

where $A_n^{HA}$ is the amplitude of the radial field for the $n$th state, $R_{n,l}(r)$ is the radial factor in the hydrogen-like wave function and $\mathcal{N}_n$ is its normalization constant. By (85), one can deduce the velocity of the PF system:

$$\dot{q}_{n,l,m_l} = r\dot{\theta}\left(1 + \frac{1}{2\pi r^2}\chi_{n,l}^2(r) S_{l,m_l}^2(\theta)\right)^{1/2} \simeq r\dot{\theta}\left(1 + \frac{1}{4\pi r^2}\chi_{n,l}^2(r) S_{l,m_l}^2(\theta)\right) \tag{87}$$

---

[8]Whenever the particle is subjected to a conservative force, the energy of the whole PF system will be conserved too. This is because, when the energy of particle is conserved, its associated field also experiences a conservative force (see, e.g., the relation (28)). Then, the energy of the whole system should be also conserved as is illustrated in the → time-independent Schrödinger equation.

where in second line, we have used the same approximation as before (see, e.g., the relation (50)). From (87), one immediately concludes that for all $s$-states ($l = 0$), $\dot{q}_{\alpha,ns} = \alpha\dot{\theta} = v_\alpha$, where $\alpha$ denotes $x$, $y$, and $z$. Writing first the components of $v$ as (80), and then integrating over $\theta$, one obtains:

$$q_{x,ns} = r\sin\theta\cos\varphi = x; \quad q_{y,ns} = r\sin\theta\sin\varphi = y; \quad q_{z,ns} = r\cos\theta = z \tag{88}$$

In other words, in all $s$-states, the electron-field orbits are the same as the classical orbits of the electron alone. That is:

$$q_{ns} = \left(q_{x,ns}^2 + q_{y,ns}^2 + q_{z,ns}^2\right)^{1/2} = r \tag{89}$$

Now, let us derive the trajectory of the PF system for $2p$-states. Using the relation (87) for $n = 2$, $l = 1$ and $m_l = 0, \pm 1$, where $S_{1,0}(\theta) = \frac{1}{2}\sqrt{6}\cos\theta$, $S_{1,\pm 1}(\theta) = \frac{1}{2}\sqrt{3}\sin\theta$ and $\chi_{2,1}(r) = A_{21}^{HA}\, r\exp(-\frac{Zr}{2a_0})$, we obtain:

$$\dot{q}_{2p_0} \simeq v\left(1 + \frac{3}{8\pi}\left(\frac{A_{21}^{HA}}{a_0}\right)^2 e^{(-\frac{Zr}{a_0})}\sin^2\theta\right) \tag{90}$$

and

$$\dot{q}_{2p_{\pm 1}} \simeq v\left(1 + \frac{3}{16\pi}\left(\frac{A_{21}^{HA}}{a_0}\right)^2 e^{(-\frac{Zr}{a_0})}\cos^2\theta\right) \tag{91}$$

When $\theta = 0$ (i.e. in $z$-direction), $\dot{q}_{2p_0} = v$ which means that the kinetic energy of the PF system reaches its minimum value. Accordingly, the probability of finding the electron in $2p_0$ state would be maximized along the $z$-direction. Similarly, for $\theta = \frac{\pi}{2}$, $\dot{q}_{2p_{\pm 1}}$ in (91) reaches its minimum value $v$ which explains why in $2p_{\pm 1}$ states, the probability of finding the electron would be maximized in $xy$-plane.

For $2p_0$ state, it is straightforward to get the $x$, $y$, and $z$-components of $q_{2p_0}$ from (90). With a little manipulation, one can derive:

$$q_{x,2p_0} \simeq x\left(1 + \frac{1}{8\pi}\left(\frac{A_{21}^{HA}}{a_0}\right)^2 e^{(-\frac{Zr}{a_0})}\sin^2\theta\right) \tag{92}$$

where $\int v_x dt = \int r\dot{\theta}\cos\theta\cos\varphi dt = r\cos\varphi \int \cos\theta d\theta = x$. Also, $\int v_x \sin^2\theta dt = \frac{1}{3}x\sin^2\theta$. Similarly, one can show that

$$q_{y,2p_0} \simeq y\left(1 + \frac{1}{8\pi}\left(\frac{A_{21}^{HA}}{a_0}\right)^2 e^{(-\frac{Zr}{a_0})}\sin^2\theta\right) \tag{93}$$

and

$$q_{z,2p_0} \simeq z\left[1 + \frac{1}{8\pi}\left(\frac{A_{21}^{HA}}{a_0}\right)^2 e^{(-\frac{Zr}{a_0})}(2 + \sin^2\theta)\right] \qquad (94)$$

Considering that $q_{2p_0} = \left(q_{x,2p_0}^2 + q_{y,2p_0}^2 + q_{z,2p_0}^2\right)^{1/2}$, and keeping only the terms including the second power of amplitude (i.e., $(A_{21}^{HA})^2$), one gets:

$$q_{2p_0} \simeq r\left[1 + \frac{1}{8\pi}\left(\frac{A_{21}^{HA}}{a_0}\right)^2 e^{(-\frac{Zr}{a_0})}(1 + \cos^2\theta)\right] \qquad (95)$$

In a similar manner, we obtain:

$$q_{2p_{\pm 1}} \simeq r\left[1 + \frac{1}{16\pi}\left(\frac{A_{21}^{HA}}{a_0}\right)^2 e^{(-\frac{Zr}{a_0})}(1 + \sin^2\theta)\right] \qquad (96)$$

From the relations (95) and (96), it is now apparent that there is a departure from a perfect circular motion in $2p$-orbitals. The amount of departure depends on the value of $\left(\frac{A_{21}^{HA}}{a_0}\right)^2$, however. Let us consider the case $\dot{q}_{2p_0}$ in (90), for instance. Keeping validate the approximation used in (87), it is sufficient to show that for the maximum value of the factor $e^{(-\frac{Zr}{a_0})}\sin^2\theta$ in (90) (that is, $e^{(-\frac{Zr}{a_0})}\sin^2\theta|_{\max} = 1$), the difference between two expressions $\left(1 + \frac{3}{8\pi}\left(\frac{A_{21}^{HA}}{a_0}\right)^2\right)$ and $\left(1 + \frac{3}{4\pi}\left(\frac{A_{21}^{HA}}{a_0}\right)^2\right)^{1/2}$ is negligible for appropriate values of $A_{21}^{HA}$. Choosing $0 < A_{21}^{HA} \leq 10^{-1}$ m, the difference is $\leq 7.1 \times 10^{-7}$ which the upper limit is sufficiently small. The same range of values for $A_{21}^{HA}$ can be chosen for $\dot{q}_{2p_{\pm 1}}$ in (91) with even better agreement.

In all above relations, depending on a definite value of $\dot{\theta}$, $r$ is also determined definitely. However, since the electron might possess a wide range of angular velocities, its distance from the nucleus could vary correspondingly. Taking into account the relation $\dot{\theta}^2 = \frac{Ze'^2}{\mu r^3}$, great values of $\dot{\theta}$ correspond to small radii and vice versa. Yet, to draw the trajectories (95) and (96), a given separation $r$ should be considered.

The plots of $\frac{q_{2p_0}}{r}$ and $\frac{q_{2p_{\pm 1}}}{r}$ are drawn in figures 2-a and b, respectively, for $r = a_0$, $Z = 1$ and $A_{21}^{HA} \approx 0.1$ m. One can see that even for low energy states, the orbits resemble the classical ones.

In a similar way, one can obtain the orbits of electron in higher states. For a given $n$, however, as $r$ increases, all $q$-functions approaches $r$ which is not a probable occasion, when electron has low energy values. It is anticipated also that the amplitude factors $A_{n,l}^{HA}$ become smaller and smaller as $n$ grows. Nevertheless, as one gives energy to prepare an excited atom, the classical energy $E_\mu$ will also increases (becomes less negative) in (77) concurrently. This in turn means that electron becomes gradually free of the Coulomb's attraction of nucleus and at the same time, the energy of field comes near to zero, as both $E_n$ and $E_\mu$ approach zero in (82). Hence, we will have a particle free of any attraction of nucleus.

# 4 Nonlinear Schrödinger equation

We saw before in section 2 that for real stationary fields, there exists an oscillating-like term in the force expression.[9] This term manifests itself in Newton-like dynamics of the one-dimensional as well as the three-dimensional stationary fields depicted by relations (9) and (28), respectively. Yet, one can assess the subject differently. The existence of a field isomorphic to the homogeneous transverse waves can shed light on the physical origin of the time-independent Schrödinger equation, so that it enables us to search for effects in the micro-physics realm beyond the predictions of standard quantum mechanics.

The procedure is as follows. Considering the relation (9), let us assume that for a one-particle one-dimensional system, there is a generalized associated stationary field $\tilde{\chi}(x(t))$ subjected to an anhormonic force described as

$$f_{F1} = -m\bar{\omega}^2 \tilde{\chi} + m\varepsilon' \tilde{\chi}^3 + \ldots \tag{97}$$

where $\varepsilon'$ is a small constant and the subscript $F1$ denotes that only the first part of field's force $f_F$ in (9) is considered here. On the other hand, from (5) one can see that $f_{F1}$ should be equal to $mv_P^2 \tilde{\chi}''$. Assuming that the two expressions are equal and considering only the first two terms in (97), we finally obtain:

$$-\frac{\hbar^2}{2m} \tilde{\chi}'' + V_P(x)\tilde{\chi} = E\tilde{\chi} - \frac{\varepsilon' \hbar^2}{2p_P^2} \tilde{\chi}^3 \tag{98}$$

where $p_P = \sqrt{2m(E_P - V_P(x))}$ is the linear momentum of particle. In the limit $\varepsilon \to 0$, the linear Schrödinger equation comes into view. Hence, we are interested to the solutions of kind

$$\tilde{\chi} = \chi + \varepsilon \chi^{(1)} + O(\varepsilon^2) \tag{99}$$

where $\chi$ is the solution of the linear time-independent Schrödinger equation and $\chi^{(1)}$ is the first order correction to $\chi$.

For a two-particle three-dimensional system exposed to a central potential $V_P = V_P(r)$, using the relation (28), one can generalize the equation (98) to:

$$-\frac{\hbar^2}{2\mu} \nabla^2 \tilde{\chi}(x,y,z) + V_P(x,y,z)\tilde{\chi}(x,y,z) = E\tilde{\chi}(x,y,z) - \frac{\varepsilon' \hbar^2}{2p_\mu^2} \tilde{\chi}^3(x,y,z) \tag{100}$$

where $\mu$ is the reduced mass and $p_\mu = \mu v_\mu$ is the linear momentum in relative coordinates. This equation can be also rewritten as

$$\nabla^2 \tilde{\chi} = -k^2 \tilde{\chi} + \frac{\varepsilon'}{\mu v_\mu^2} \tilde{\chi}^3 \tag{101}$$

---

[9] In fact, usually, the stationary fields are either real functions or they can be made real. An exception, however, is free particle's state. Yet, for a free particle, one can deduce from the relation (5) that $f_F = 0$. So, this case does not concern what we discuss in this section.

where $k^2 = \frac{2\mu}{\hbar^2}(E - V_P(r))$. This is a Duffing-like equation in which the $k$-factor is not generally constant. Correspondingly, the time-dependent Schrödinger equation can be presented as:

$$i\hbar \frac{\partial}{\partial t} \tilde{X}(\vec{r}, t) = \frac{-\hbar^2}{2\mu} \nabla^2 \tilde{X}(\vec{r}, t) + V_P \tilde{X}(\vec{r}, t) + \frac{\varepsilon' \hbar^2}{2p_\mu^2} \tilde{X}^3(\vec{r}, t) \tag{102}$$

Interestingly, the dual correspondence between a particle's field $\tilde{\chi}$ and its wave function goes out of sight under the nonlinear regime. To have a relation like (8), we should define a normalization constant $\mathcal{N}$ which is obtained by normalizing the probability density functions $|\psi|^2$. However, if one considers $|\psi|^2 \propto |\tilde{\chi}|^2$, he or she will be forced to adopt the same density function achieved in linear condition, since the nonlinear terms in $|\tilde{\chi}|^2$ (including higher powers of amplitude $A$ as well as $\varepsilon'$) are so small that can be definitely ignored. So, as far as one intends to evaluate the probability of events, linear equations are sufficient. In other words, in (99) we can assume that $\tilde{\chi} \simeq \chi$. The equations (101) and (102) give us a more accurate form of fields, but this has no primary importance. What is more important is that we can obtain discernible fine structures of the energy levels not predicted by the ordinary Schrödinger equation or we are able to describe time-dependent phenomena in which nonlinear effects have discernible manifestations.

As an instance, we solve equation (101) for the case of a particle in one-dimensional box examined before in section 3.1 using linear Schrödinger equation. Here, we should solve the following equation:

$$\tilde{\chi}''(u) + \tilde{\chi}(u) - \frac{\varepsilon}{k^2} \tilde{\chi}^3(u) = 0 \tag{103}$$

where $u = kx$ and $0 \leq x \leq a$, $k = \frac{2mE}{\hbar^2}$ and $\varepsilon = \frac{\varepsilon'}{mv_P^2}$. Using the method of scaled parameters, one can find the solution of (103) as [8]:

$$\tilde{\chi}(x) = \tilde{A}\cos\left[\left(1 - \frac{3\varepsilon}{8k^2}\tilde{A}^2\right)kx + B\right] - \frac{1\varepsilon}{32k^2}\tilde{A}^3 \cos\left[3\left(1 - \frac{3\varepsilon}{8k^2}\tilde{A}^2\right)kx + 3B\right] + O(\varepsilon^2) \tag{104}$$

where $\tilde{A}$ is a constant with dimension of length. The boundary condition $\tilde{\chi}(0) = \tilde{\chi}''(0) = 0$ leads to the conclusion that $B = \pm \frac{\pi}{2}$. From the other boundary condition $\tilde{\chi}(a) = \tilde{\chi}''(a) = 0$, one can infer that

$$\left(1 - \frac{3\varepsilon}{8k^2}\tilde{A}_n^2\right)k_n a = \pm n\pi \tag{105}$$

From (105) with some manipulation, one can show that up to the first order of $\varepsilon$,

$$k_n^2 = \frac{n^2\pi^2}{4a^2}\left[1 + \left(1 + \frac{3}{2}\frac{\varepsilon \tilde{A}_n^2 a^2}{n^2\pi^2}\right)^{1/2}\right]^2 \tag{106}$$

Inserting the relation (105) in (104) we get:

$$\tilde{\chi}_n(x) = \tilde{A}_n \sin\left(\frac{n\pi x}{a}\right) + \frac{\varepsilon}{32 k_n^2} \tilde{A}_n^3 \sin\left(\frac{3n\pi x}{a}\right) + O(\varepsilon^2); A_n \sin\left(\frac{n\pi x}{a}\right) \tag{107}$$

where we have assumed that the factor $\frac{\varepsilon}{32 k_n^2}\tilde{A}_n^3$ is sufficiently small to be neglected. The coefficient $A_n$ has been defined in (47).

From (106), we can obtain a relation for the transitional energy levels of the system:

$$E_n \simeq \frac{n^2 h^2}{8ma^2}\left[\frac{1}{2} + \left(\frac{1}{4} + \frac{3}{8}\frac{\varepsilon A_n^2 a^2}{n^2 \pi^2}\right)^{1/2}\right]^2 + O(\varepsilon^2) \tag{108}$$

where we have estimated $\tilde{A}_n \simeq A_n$. If $\varepsilon \to 0$, $E_n \to \frac{n^2 h^2}{8ma^2}$. Taking into account the relations (107) and (108), we emphasize again that nonlinear effects are negligible in our approach, when fields and wave functions are purposed, but are important for discernible quantities. Any experimental verification of (108) confirms the existence of the energetic fields, due to the presence of both $\varepsilon$ (an indication of the mechanical-like aspects of field) and $A_n$ factors.

## 5 Conclusion

Supposing a particle and its probability field as an unified objective entity in micro-world which its dynamics is governed by a Newtonian-like equation, we showed how quantum dynamics could be reconciled with classical rules in a subtle way. We elaborated the nature of stationary fields, providing a new scheme for analyzing different features of a PF system affected by a conservative force. Due to the real existence of particle, its associated field is endowed with some mechanical-like attributes which enables us to develop the theory to provide new predictions including nonlinear effects at a more subtle level of observation.

Regarding a system composed of many particles, we argued how the physical status of each individual PF system could be influenced by the whole character of the total system in a local context-dependent manner. Hence, we have developed the essence of a contextual causal theory in which each particle has an associated probability field which its existence in micro-domain is an inseparable ingredient of the physical reality. Based on the local causal characteristic of this theory as well as its capability to be reconsidered on a geometric basis (see the relation (11)), its development to a more fundamental theory comprising -both general and special- relativistic effects should be seriously thought about.

**Acknowledgment**

The author would like to extend his sincere thanks to Abouzar Massoudi, Mohammad Bahrami, Majid Karimi and Iman Khatam for their encouragement and assistance in preparing this article. Also, I would like to thank my students Zahra Mashhadi, Farhad T. Ghahramani, Arash Tirandaz, Hossein Sadeghi, Keivan Sadri and Shobeir K. Seddigh for the stimulation and suggestions I have received from them.

**Appendix A**

Considering the time-independent Schrödinger equation $\psi'' = -\frac{p^2}{\hbar^2}\psi$ and multiplying both sides by $\psi^*$, it is straightforward to show that the average of $p^2$ (denoted by $\langle p^2 \rangle$) is equal to:

$$\begin{aligned}\langle p^2 \rangle &= \int_{\Omega(x)} dx\ p^2 |\psi|^2 = \int_X dx\ \psi^* p^2 \psi \\ &= -\hbar^2 \int_{\Omega(x)} dx\ \psi^* \psi'' = \int_{\Omega(x)} dx\ \psi^* (-\hbar^2 \psi'')\end{aligned} \qquad \text{(A-1)}$$

From this relation, one can define the following operator for $p^2$ (denoted by $\hat{p}^2$):

$$\hat{p}^2 := -\hbar^2 \psi'' = (\pm i\hbar \tfrac{\partial}{\partial x})(\pm i\hbar \tfrac{\partial}{\partial x}) = \hat{p}\hat{p} \qquad \text{(A-2)}$$

For the reason of consistency, we then define the operator of $p$ as

$$\hat{p} := (-i\hbar \tfrac{\partial}{\partial x}) \qquad \text{(A-3)}$$

which is a linear Hermitian operator.[10] The procedure used here can be generalized to be applicable for any observable in the micro-physical domain. Hence, we assume that to every physical observable there corresponds a linear Hermitian operator which can be defined by using the linear Hermitian operators $x$ (denoted by $\hat{x}$) and the $x$ component of momentum $p_x$ (denoted by $\hat{p}_x = -i\hbar \frac{\partial}{\partial x}$). To find the desired operator, it is sufficient to write down the classical expression for the observable in terms of spatial coordinates and corresponding linear-momentum components. Then, one can replace each coordinate $x$ by $\hat{x}$ and each momentum component $p_x$ by $\hat{p}_x$.

**Appendix B**

Here, we are going to examine the conditions under which the approximation (50) is valid. Let us write $f_n(x)$ in (50) as

$$f_n(x) = (1 + b_n^2 \cos^2(k_n x))^{1/2} \qquad \text{(B-1)}$$

where $b_n^2 = \left(\frac{p_n^2}{p_P^2} - 1\right)$ and $k_n = \frac{n\pi}{a}$. Expanding $f_n(x)$ in terms of $\chi'_n = b_n \cos(k_n x)$, we obtain:

$$f_n(x) = 1 + \frac{b_n^2 \cos^2(k_n x)}{2} - \frac{b_n^4 \cos^4(k_n x)}{8} + \frac{b_n^6 \cos^6(k_n x)}{16} - \frac{5 b_n^8 \cos^8(k_n x)}{128} + \ldots \qquad \text{(B-2)}$$

For the series converges, it is necessary that $0 \leq b_n^2 < 1$, which means that $0.5 < \frac{p_P^2}{p_n^2} \leq 1$. This is in accordance with our earlier assumption that field's energy is not so large relative to the energy of particle.

---

[10] By the term Hermitian, we mean that for every physical observable $B$, we should have $\langle B \rangle = \langle B \rangle^*$.

Using (49) and (B-2), one can derive the following relation for $q_n(x)$:

$$q_n(x) = g_{n,PF}\left[b_n^{(1)}x + \frac{b_n^{(2)}}{k_n}\sin(2k_n x) - \frac{b_n^{(3)}}{k_n}\sin(4k_n x) + \ldots\right] \quad \text{(B-3)}$$

Here, we have:

$$b_n^{(1)} = 1 + \frac{b_n^2}{4} - \frac{3b_n^4}{64} + \frac{5b_n^6}{256} - \frac{175b_n^8}{16384}; \quad \text{(B-4)}$$

$$b_n^{(2)} = \frac{b_n^2}{8} - \frac{b_n^4}{32} + \frac{15b_n^6}{1024} - \frac{35b_n^8}{4096}; \quad \text{(B-5)}$$

and

$$b_n^{(3)} = \frac{b_n^4}{256} - \frac{3b_n^6}{1024} + \frac{35b_n^8}{16384} \quad \text{(B-6)}$$

Assuming that $b_n^2 = 0.5$ (or, $\frac{p_p^2}{p_n^2} = \frac{2}{3}$) for a definite $n = n^*$ (e.g., $n^* = 1$), for the maximum value of $\cos^2(k_{n^*}x)$, (i.e., when $\cos^2(k_{n^*}x) = 1$), one obtains $f_{n^*} \simeq 1.224$ up to the eighth order of $b_n$ in (B-2), while the exact value of $f_{n^*}$ is about $1.225$ in (B-1). The difference is about $10^{-3}$ which shows $f_n$ in (B-2) is a good candidate for getting a more general form of $q_n(x)$ in (B-3).

Considering the same value of $b_{n^*}^2 = 0.5$ in (B-3) and using the relations (B-4), (B-5) and (B-6) we obtain: $b_{n^*}^{(1)} \approx 1.115$, $b_{n^*}^{(2)} \approx 0.056$ and $b_{n^*}^{(3)} \approx 7.439 \times 10^{-4}$ which the last term is small enough to be ignored. Defining $g_{n^*,PF} = \frac{1}{b_{n^*}^{(1)}}$ in (B-3), we get:

$$q_{n^*}(x) \simeq x + \frac{0.05}{k_{n^*}}\sin(2k_{n^*}x) \quad \text{(B-7)}$$

For $b_{n^*}^2 = 0.5$, the r.h.s. of equation (52) is approximately equal to $x + \frac{0.056}{k_{n^*}}\sin(2k_{n^*}x)$ which shows the difference between the second coefficient in (52) and (B-7) is about $6 \times 10^{-3}$. For smaller values of $b_n^2$, this difference becomes even smaller, but for $0.5 \leq b_n^2 < 1$ it is better to consider the third term in (B-3) too. However, under the latter regime, the energy of field also increases, which is not supported from a physical point of view.

**Appendix C**

Here, we first derive the relation (65). According to relation (64), for an oscillating PF system in the ground energy state we have

$$q_0^{osc}(\bar{r}) \simeq \bar{r} + \frac{(\alpha A_0^{osc})^2}{16\pi}\int d\bar{r}\ \bar{r}^2 e^{-\alpha\bar{r}^2} + \ldots \quad \text{(C-1)}$$

Expanding the exponential term as $e^{-\alpha\bar{r}^2} = \sum_{k=0}\frac{(\alpha\bar{r}^2)^k}{k!}$, we get

$$q_0^{osc}(\bar{r}) \simeq \bar{r} + \frac{(\alpha A_0^{osc})^2 \bar{r}^3}{48\pi}\left(1 - \frac{3\alpha \bar{r}^2}{5} + \frac{3\alpha^2 \bar{r}^4}{14} + \ldots\right)$$

$$\simeq \bar{r} + \frac{(\alpha A_0^{osc})^2 \bar{r}^3}{48\pi}\left[1 - \alpha \bar{r}^2 + \left(\frac{2\alpha \bar{r}^2}{5}\right) + \frac{\alpha^2 \bar{r}^4}{2} - \left(\frac{2\alpha^2 \bar{r}^4}{7}\right) + \ldots\right]$$

$$\simeq \bar{r} + \frac{(\alpha A_0^{osc})^2 \bar{r}^3}{48\pi}\left\{e^{-\alpha \bar{r}^2} + \frac{2\alpha \bar{r}^2}{5}\left[(1 - \alpha \bar{r}^2 + \ldots) + \left(\frac{2\alpha \bar{r}^2}{7}\right) + \ldots\right] + \ldots\right\} \quad \text{(C-2)}$$

The last relation in (C-2) can then be written as:

$$q_0^{osc}(\bar{r}) \simeq \bar{r} + \frac{(\alpha A_0^{osc})^2 \bar{r}^3}{48\pi} e^{-\alpha \bar{r}^2} + \frac{\alpha^3 (A_0^{osc})^2 \bar{r}^5}{120\pi} e^{-\alpha \bar{r}^2} + \ldots \quad \text{(C-3)}$$

In a similar manner, for $n = 1$, we have

$$q_1^{osc}(\bar{r}) \simeq \bar{r} + \frac{\alpha (A_1^{osc})^2}{16\pi} \int d\bar{r}\, (1 - \alpha \bar{r}^2)^2 e^{-\alpha \bar{r}^2} + \ldots \quad \text{(C-4)}$$

which can be written as:

$$q_1^{osc}(\bar{r}) \simeq \bar{r} + \frac{\alpha (A_1^{osc})^2}{16\pi} \int d\bar{r}\, \left(1 - 3\alpha \bar{r}^2 + \frac{7\alpha^2 \bar{r}^4}{2} - \frac{13\alpha^3 \bar{r}^6}{6} + \ldots\right)$$

$$\simeq \bar{r} + \frac{\alpha (A_1^{osc})^2 \bar{r}}{16\pi}\left[1 - \alpha \bar{r}^2 + \frac{\alpha^2 \bar{r}^4}{2} + \left(\frac{\alpha^2 \bar{r}^4}{5}\right) - \frac{\alpha^3 \bar{r}^6}{6} - \left(\frac{\alpha^3 \bar{r}^6}{7}\right) + \ldots\right]$$

$$\simeq \bar{r} + \frac{\alpha (A_1^{osc})^2 \bar{r}}{16\pi}\left\{e^{-\alpha \bar{r}^2} + \frac{\alpha^2 \bar{r}^4}{5}\left[(1 - \alpha \bar{r}^2 + \ldots) + \left(\frac{2\alpha \bar{r}^2}{7}\right) + \ldots\right] + \ldots\right\} \quad \text{(C-5)}$$

Accordingly, one can show that:

$$q_1^{osc}(\bar{r}) \simeq \bar{r} + \frac{\alpha (A_1^{osc})^2 \bar{r}}{16\pi} e^{-\alpha \bar{r}^2} + \frac{\alpha^3 (A_1^{osc})^2 \bar{r}^5}{80\pi} e^{-\alpha \bar{r}^2} + \ldots \quad \text{(C-6)}$$

Figure Captions

Figure Caption 1 The trajectories of a PF system in one dimensional box with width $a = 2$ nm are plotted in comparison with a dashed straight line for (a) $q_1(x)$ with $\frac{p_1^2}{p_P^2} = 1.5$, (b) $q_2(x)$ with $\frac{p_2^2}{p_P^2} = 1.45$, and (c) $q_3(x)$ with $\frac{p_3^2}{p_P^2} = 1.40$.

Figure Caption 2 (a) Graph of $\frac{q_{2p_0}}{r}$ for a hydrogen atom with $r = a_0$, $Z = 1$ and $A_{21}^{HA} \approx 0.1$ m. The lengths of the major (minor) diameters along the vertical (horizontal) axis are 1.00029 and 1.00015 rad, respectively. (b) Graph of $\frac{q_{2p_{\pm 1}}}{r}$ with the same values of $r$, $Z$ and $A_{21}^{HA}$ as (a). The lengths of the major (minor) diameters along the vertical (horizontal) axis are, respectively, 1.00015 and 1.000073 rad here.

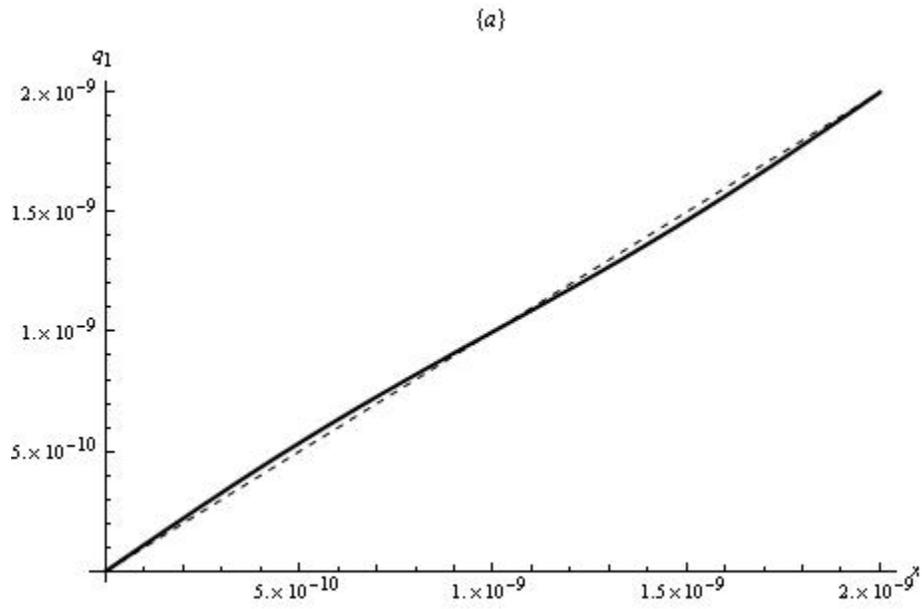

Figure 1(a)

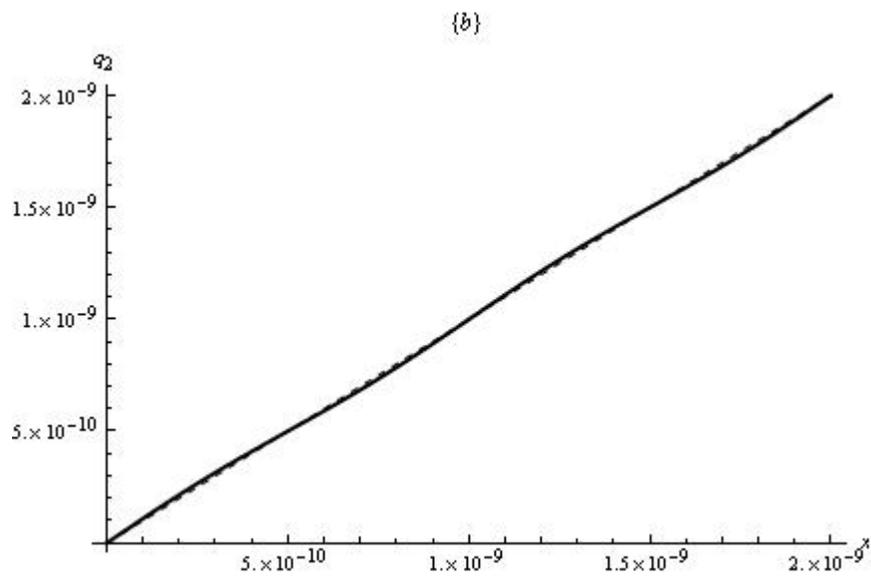

Figure 1(b)

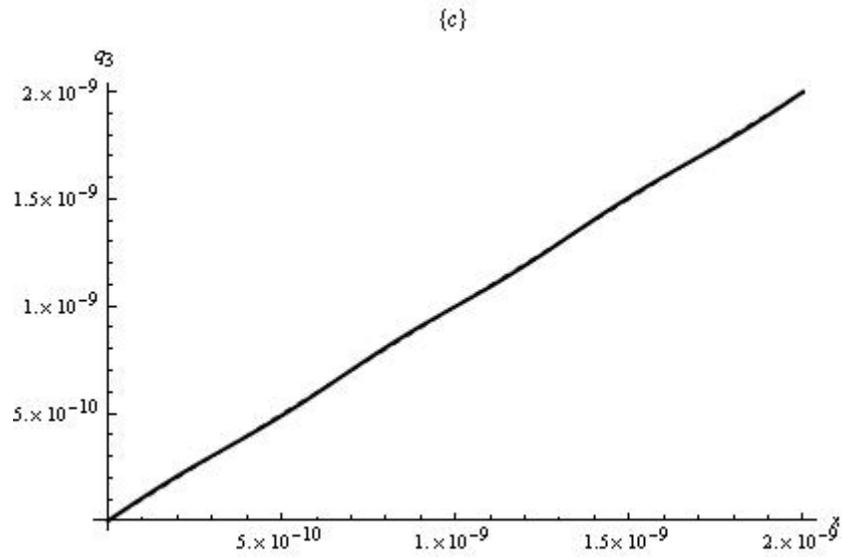

Figure 1(c)

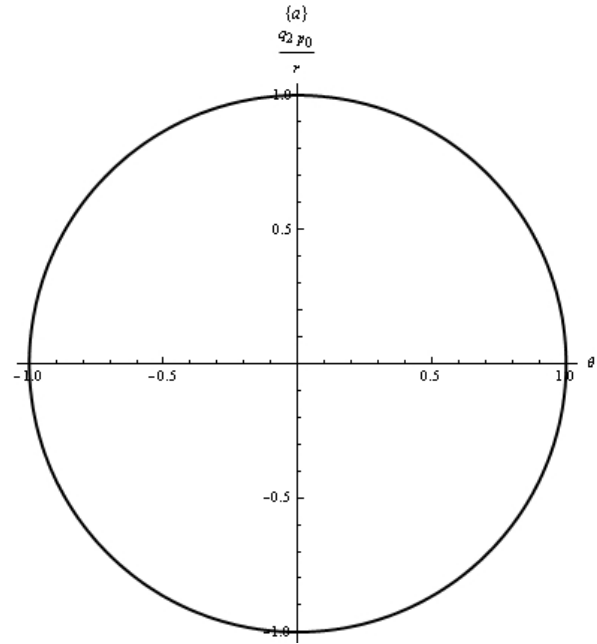

Figure 2(a)

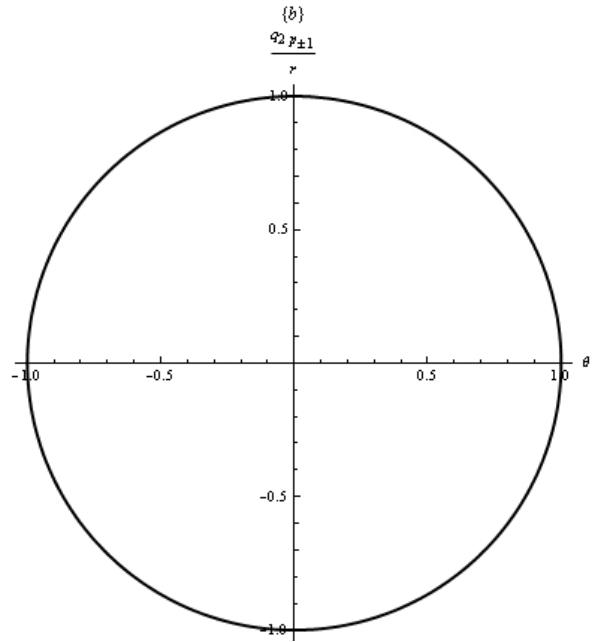

Figure 2(b)